\documentclass[
    ,final            
  ]
  {aipproc}

\layoutstyle{8x11single}

\begin{document}

\title{\vspace{-2.1cm}
\hfill{\hspace{13cm}\small{\textnormal{CPHT-PC050.0513 }}}\\
\vspace{-0.3cm}\hfill{\hspace{13cm}\small{\textnormal{DESY 13-095}}}\\[1.065cm]Hidden Photons in connection to Dark Matter}


\author{Sarah Andreas}{
  address={Deutsches Elektronen-Synchrotron DESY, Hamburg, Germany}
}

\author{Mark D.\ Goodsell}{
  address={CPhT, Ecole Polytechnique, Palaiseau, France}
}

\author{Andreas Ringwald}{
  address={Deutsches Elektronen-Synchrotron DESY, Hamburg, Germany}
}

\begin{abstract}
Light extra U(1) gauge bosons, so called hidden photons, which reside in a hidden sector have attracted much attention since they are a well motivated feature of many scenarios beyond the Standard Model and furthermore could mediate the interaction with hidden sector dark matter. We review limits on hidden photons from past electron beam dump experiments including two new limits from such experiments at KEK and Orsay. In addition, we study the possibility of having dark matter in the hidden sector. A simple toy model and different supersymmetric realisations are shown to provide viable dark matter candidates in the hidden sector that are in agreement with recent direct detection limits.
\end{abstract}

\maketitle



There are various motivations both from a theoretical and a phenomenological point of view for gauge bosons of an extra U(1) symmetry in a hidden sector. These new gauge bosons are called hidden photons. They arise naturally in scenarios beyond the Standard Model (SM) like string theory and supersymmetry. Furthermore, the hidden photon is of interest since it provides a solution to the discrepancy in the muon anomalous magnetic moment between the SM prediction and the experimentally measured value~\cite{Pospelov:2008zw}. In addition, models that contain a dark matter (DM) candidate in the hidden sector which interacts with the SM through a light hidden photon attracted much attention in the context of recent astrophysical observations~\cite{Fayet:2007ua:ArkaniHamed:2008qn:Cheung:2009qd,Pospelov:2007mp:Chun:2010ve:Mambrini:2011dw:Hooper:2012cw}.

The dominant interaction at low energies between the hidden photon $\gamma'$ and the visible sector is through kinetic mixing with the ordinary photon. Generated, for example, by loops of heavy particles charged under both U(1)s~\cite{Holdom:1985ag}, a reasonable estimate for the size of the kinetic mixing is of the order of a loop factor $\mathcal{O}(10^{-3}-10^{-4})$ from integrating out those particles. We therefore impose a relation between the hidden gauge coupling $g_h$ and the kinetic mixing as
\begin{equation}
\chi_Y = \frac{g_Y g_h}{16 \pi^2} \ \kappa \, ,\label{eq-kappa}
\end{equation}
with $\kappa$ being $\sim\mathcal{O}(1)$ and depending logarithmically on the masses of the particles in the loop. In supersymmetric theories, this relation is also enforced by holomorphy.

For the most simple hidden sector with an extra U(1) symmetry and the hidden photon $\gamma'$, the low energy effective Lagrangian describing the kinetic mixing with the photon at masses much below the $Z$ mass is given by
\begin{equation}
{\cal L} = -\frac{1}{4}F_{\mu \nu}F^{\mu \nu} - \frac{1}{4}X_{\mu \nu}X^{\mu \nu}
+ \frac{\chi}{2} X_{\mu \nu} F^{\mu \nu} + \frac{m_{\gamma^{\prime}}^2}{2}  X_{\mu}X^\mu + e j^\mu_{\mathrm{em}} A_\mu \, ,
\end{equation}
where $\chi \equiv \chi_Y c_W$ with the cosine $c_W$ of the Weinberg angle, $F_{\mu\nu}$ and $X_{\mu\nu}$ are the field strengths and $A_\mu$ and $X_\mu$ the gauge fields of the ordinary and hidden U(1), respectively. The kinetic mixing with the photon generates an effective coupling of the hidden photon with the electromagnetic current $j^\mu_{\mathrm{em}}$ of the SM and allows the hidden photon to be probed experimentally. The mass $m_{\gamma'}$ of the hidden photon can result either from the Higgs or St\"{u}ckelberg mechanism. Masses around the GeV scale are obtained naturally in both cases but also much smaller values are possible~\cite{Goodsell:2009xc:Cicoli:2011yh}. For these GeV-scale hidden photons, an important class of constraints arises from electron beam dump experiments. Limits from past beam dump experiments at SLAC and Fermilab were studied in~\cite{Bjorken:2009mm} and further in~\cite{Andreas:2012mt} which additionally considered two other experiments at KEK and in Orsay and took the experimental acceptances into account. An overview of all current experimental constraints on the hidden photon for the MeV to GeV mass range is given in~\cite{Andreas:2012mt,Andreas:2012xh:Hewett:2012ns:Andreas:2011xf}. In general, hidden sectors can not only contain gauge but also matter fields. In particular, in various models the interaction of DM in the hidden sector through a GeV-scale hidden photon has been studied~\cite{Fayet:2007ua:ArkaniHamed:2008qn:Cheung:2009qd,Pospelov:2007mp:Chun:2010ve:Mambrini:2011dw:Hooper:2012cw,Andreas:2011in,GeVscale}. 

This paper summarises the current status of limits on hidden photons from electron beam dump experiments, based on the results presented in~\cite{Andreas:2012mt}. Furthermore, a toy-model as well as different supersymmetric models for hidden sectors with DM interacting via a hidden photon are discussed regarding the DM relic density and constraints from recent direct detection experiments, following and updating the analysis of~\cite{Andreas:2011in}.

\section{Hidden photons in electron beam dump experiments} \label{sec-HP}

\begin{figure}[htb!]
\centering{\hspace{-0.1cm}
\includegraphics[height=5.45cm]{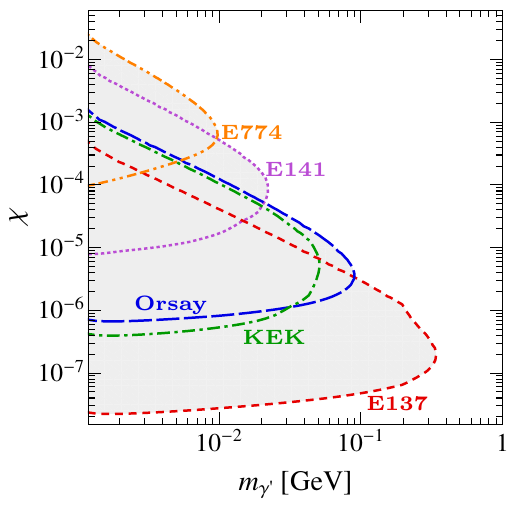}
\raisebox{3pt}{\includegraphics[height=5.42cm]{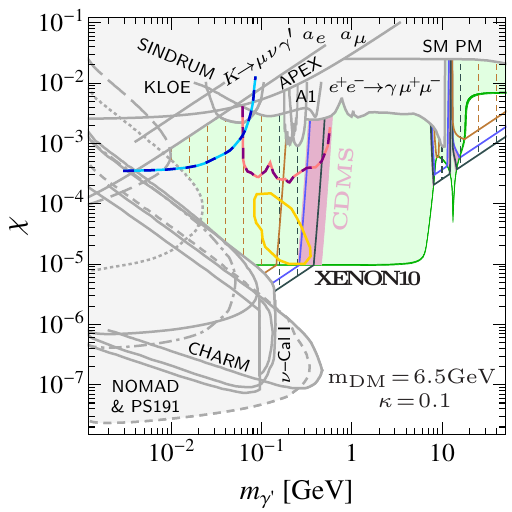}}
\raisebox{0.pt}{\includegraphics[height=5.4522cm]{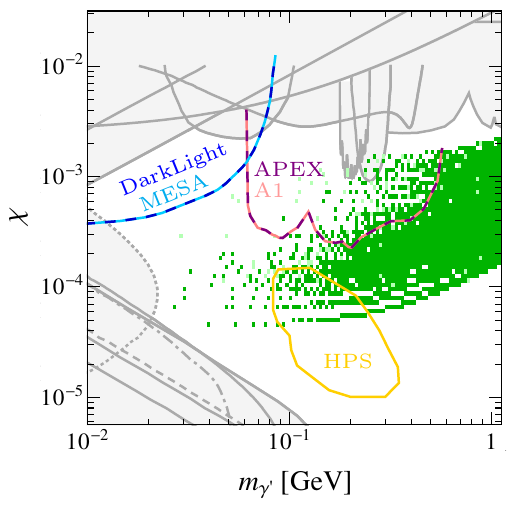}}
\caption{$\!\!$\textit{Left:} Limits on hidden photons from electron beam dumps at SLAC (E141, E137), Fermilab (E774), KEK and Orsay.\newline
\textit{Middle:} DM toy model with updated DD limits from XENON10 (black), XENON100 (blue), DAMIC (brown) and the CDMS preferred band (magenta). Dark (light) green regions give a correct (subdominant) relic abundance. Grey excluded areas as in~\cite{Andreas:2012mt,Andreas:2012xh:Hewett:2012ns:Andreas:2011xf}. \newline
\textit{Right:} Viable models with visible sector induced breaking and light hidden photon. Coloured lines for future searches as in~\cite{Andreas:2012xh:Hewett:2012ns:Andreas:2011xf}.} \label{fig-HPbeamdump}}
\end{figure}

Hidden photons can be produced in a process similar to ordinary bremsstrahlung by an electron beam incident on a target. The production cross section scales as $\sigma_{\gamma'} \propto \alpha^3 Z^2 (\chi/m_{\gamma'})^2$ and is $\mathcal{O}(10\ \mathrm{pb})$ for typical values. Carrying most of the initial beam energy, the hidden photons are highly boosted and emitted at small angles in forward direction. Due to the weak interactions with the SM the hidden photons traverse the dump and can be observed in a detector through their decay into leptons. The number of events of, e.g., $e^+ e^-$ from hidden photons decaying behind the dump and before the detector has been estimated in~\cite{Bjorken:2009mm,Andreas:2012mt}, where detailed calculations and expressions are given. Comparing this predicted number of events with the observed one allows us to derive exclusion limits for different experiments. These limits also include the experimental acceptances in order to account for the geometry and finite size of the detector as well as possible energy cuts applied in the data analysis. The acceptances are obtained by analysing simulations, preformed with the MadGraph Monte Carlo generator, for the bremsstrahlung production and the subsequent decay of hidden photons in view of the specific experimental set-ups, cf.~\cite{Andreas:2012mt}. With these acceptances, we derive new limits for experiments at KEK in Japan~\cite{Konaka:1986cb} and at Orsay in France~\cite{Davier:1989wz} and additionally reanalyse the limits presented in~\cite{Bjorken:2009mm} from E141 and E137 at SLAC as well as E774 at Fermilab. All our limits are shown in Fig.~\ref{fig-HPbeamdump} (\textit{left}). This type of experiment can not probe the upper right corner of the parameter space since there the decay length $l_{\gamma'} \propto  E_{\gamma'}/ (\chi m_{\gamma'})^2$ is so short that the hidden photons decay inside the dump and can not be observed. At small values of $\chi$, the reach of the experiments is limited by statistics and the exclusion lines scale roughly as $\propto \chi^4 L_\mathrm{dec}$, independently of $m_{\gamma'}$. Further limits and sensitivities of future searches for hidden photons are shown in Fig.~\ref{fig-HPbeamdump} (\textit{middle \& right}), see~\cite{Andreas:2012mt,Andreas:2012xh:Hewett:2012ns:Andreas:2011xf} for details.

\section{Hidden sectors with dark matter interacting via hidden photons}\label{sec-DM}

In the following, we consider models that contain besides the hidden photon also DM in the hidden sector. For a toy model and more complete supersymmetric realisations, we require the DM relic density to agree within $3\sigma$ with the value measured by Planck~\cite{Ade:2013lta} ($\Omega_\mathrm{DM} h^2 = 0.1196 \pm 0.0031$) and the scattering cross sections to fulfil the most recent direct detection (DD) constraints. Furthermore, we explore whether the spin-independent scattering can explain the events observed by CDMS~\cite{Agnese:2013rvf}. These results are an update of the ones in~\cite{Andreas:2011in} where more details on the models and computations are given. The DM annihilation in these models should be safe from indirect constraints especially when it proceeds via the hidden photon to four SM particles in the final state, but we will return to this in future work.

\subsection{Toy model: Dirac fermion as dark matter candidate}%

\begin{figure}[hbt!]
\centering{ 
\includegraphics[width=0.33\textwidth]{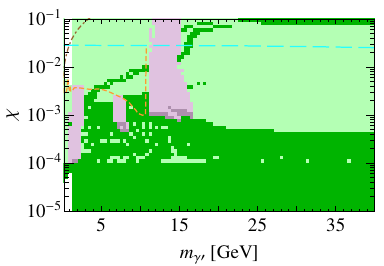}
\includegraphics[width=0.33\textwidth]{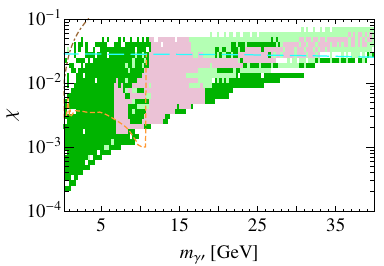}
\includegraphics[width=0.33\textwidth]{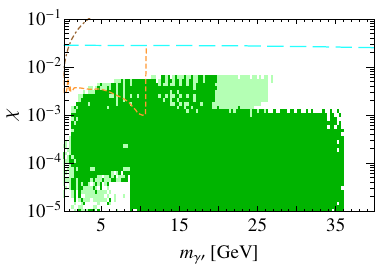}
\caption{Viable models for different hidden DM scenarios: the toy model with $\kappa=1$ (\textit{left}) and the supersymmetric models scanned over $0.1 \leq \kappa \leq 10$ with visible sector induced breaking (\textit{middle}) and with radiative breaking domination (\textit{right}). The DM particle has the correct relic abundance in the dark green regions, is subdominant in the light green ones and can explain the CDMS events in the magenta regions. Limits from SM precision measurements and the muon $g-2$ are shown as long-dashed cyan and dashed-dotted brown line, respectively, cf.~\cite{Andreas:2011in} for details. The limit from a BaBar search (short-dashed orange line) is already applied to models where the hidden photon decays entirely into SM fermions while it does not apply directly to all plotted points above this line since there the decay inside the hidden sector is possible and a model-dependent limit needs to be determined. All points are in agreement with the updated DD constraints. } \label{fig-HPDM}
}
\end{figure}

Here we consider the most simple hidden sector containing in addition to the hidden photon only one Dirac fermion as DM candidate; similar models were studied in~\cite{Pospelov:2007mp:Chun:2010ve:Mambrini:2011dw:Hooper:2012cw}. The DM annihilation proceeds either in the $s$-channel through a hidden photon into SM fermions or in the $t$-channel into two real hidden photons. Fixing the hidden gauge coupling $g_h$ via Eq.~\eqref{eq-kappa} as a function of the kinetic mixing $\chi$ and the parameter $\kappa$, the model gives the correct relic abundance for a DM mass of $6.5\ \mathrm{GeV}$ and $\kappa=0.1$ in the dark green band in Fig.~\ref{fig-HPbeamdump} (\textit{middle}) and a subdominant DM candidate in the light green area. Decreasing (increasing) $\kappa$ moves the dark green band downwards (upwards). The resonance at $m_{\gamma'} = 13\ \mathrm{GeV}$ results from the $s$-channel annihilation. The $t$-channel is kinematically only accessible for $m_\mathrm{DM}>m_{\gamma'}$ where it is dominant and roughly independent of $m_{\gamma'}$, thus leading to the horizontal green line at small $m_{\gamma'}$. For direct detection, the elastic scattering mediated by the hidden photon is spin-independent and occurs mostly on protons. Cross sections obtained with micrOMEGAs are rescaled by the DM abundance in the subdominant region and compared to the most recent DD bounds~\cite{Barreto:2011zu:Angle:2011th:Aprile:2012nq}. The strongest limits are shown in Fig.~\ref{fig-HPbeamdump} (\textit{middle}) for XENON10 (black), XENON100 (blue) and DAMIC (brown) and exclude a large region of the parameter space. A part of the magenta band in which according to~\cite{Kai,Frandsen:2013cna} the cross sections can account for the three events observed by CDMS is in agreement with all constraints. The grey shaded areas are excluded by the electron beam dump experiments discussed above and various other constraints summarised in~\cite{Andreas:2012mt,Andreas:2012xh:Hewett:2012ns:Andreas:2011xf}. A scan over the DM mass for fixed $\kappa=1$ gives the scatter plot in Fig.~\ref{fig-HPDM} (\textit{left}). Again dark green represents models which give the correct relic abundance and light green those with a subdominant DM candidate. In the magenta regions the scattering cross sections can explain the CDMS events (lighter shades for subdominant DM, darker shades for the total relic abundance). The plot only contains models that are in agreement with all updated DD limits. Thus, the Dirac fermion of the toy model is a viable DM candidate for a wide range of parameters with the total or a subdominant relic abundance and the capability to explain the CDMS events. 

\subsection{Supersymmetric model: Majorana and Dirac fermion as dark matter candidates}

To obtain the simplest anomaly-free supersymmetric realisation of a dark sector model, we consider the three chiral superfields $S$, $H_+$ and $H_-$, where $H_\pm$ have charges~$\pm 1$ under the hidden U(1), as well as the superpotential $W \supset \lambda_S S H_+ H_-$ with the Yukawa coupling $\lambda_S$. We study the DM phenomenology for two different scenarios in which the hidden gauge symmetry is broken by two different mechanisms, as described in the following and in more detail in~\cite{Andreas:2011in}. All results assume the MSSM in the visible sector but are mostly independent of this choice.

One possibility is to break the hidden gauge symmetry by the effective Fayet--Iliopoulos term induced by kinetic mixing with the visible Higgs D-term. In this case, the DM particle is either a Dirac or a Majorana fermion. As in the toy model, the Dirac fermion scatters spin-independently and is constrained by XENON10, XENON100 and DAMIC. The scattering of the Majorana fermion, due to its axial coupling, is mainly spin-dependent and limits arise from SIMPLE, PICASSO and COUPP for scattering on protons~\cite{Felizardo:2011uw:Archambault:2012pm:Behnke:2012ys} and from XENON100 for scattering on neutrons~\cite{Aprile:2013doa}. The Majorana fermion has also some spin-independent scattering arising from the Higgs-portal term, but the cross sections are well below the reach of current experiments. Applying the different constraints on both particles, the viable models are shown in the scatter plot of Fig.~\ref{fig-HPDM} (\textit{middle}) where we scanned both over the DM mass and $\kappa$ in the range $0.1\leq\kappa\leq 10$. Models in the dark green regions provide the correct relic abundance and those in the light green regions have a subdominant DM candidate. The Dirac fermion has a similar phenomenology as in the previous section and can explain the events observed by CDMS in the magenta regions of the parameter space (in contrast to the toy model, however, only as subdominant DM). Unlike in the toy model, the DM particle in the supersymmetric cases can not be heavier than the hidden photon so that the $t$-channel annihilation is not possible and there are no viable models in the lower part of the plot. A scatter plot of a scan for low hidden photon and DM masses is shown in Fig.~\ref{fig-HPbeamdump}~(\textit{right}).

In the second scenario, the hidden gauge symmetry is broken by the running of the Yukawa coupling $\lambda_S$. The DM candidate is again a Majorana fermion with spin-dependent scattering. Applying the corresponding DD constraints, the models in the scatter plot of Fig.~\ref{fig-HPDM} (\textit{right}) give the total relic abundance or a subdominant contribution as indicated in dark and light green, respectively. Here, we again scanned over the DM mass and $\kappa$ in the range $0.1\leq\kappa\leq10$.

Thus, various supersymmetric models yield viable dark sectors with either a Dirac or a Majorana fermion as DM. They can be probed by DD experiments with spin-independent or spin-dependent scattering, respectively, and give the total or a subdominant relic abundance. Furthermore, the Dirac fermion can explain the events observed by CDMS.

\section{Conclusions}
A hidden sector with a dark force is well motivated from a theoretical (string theory, SUSY) and phenomenological ($g-2$, DM) point of view. Past electron beam dump experiments provide an important class of constraints on the mass and kinetic mixing of the hidden photon. We presented different dark sector models, in particular supersymmetric ones with gravity mediation, that provide viable DM candidates with either the right or a subdominant relic abundance. The Dirac fermion DM candidate in a toy model or in supersymmetric models with visible sector induced breaking has spin-independent scattering and can explain the CDMS events while satisfying the DD limits. In these supersymmetric scenarios or other ones with radiative breaking domination, also a Majorana fermion can be obtained as DM particle and can fulfil the DD constraints on spin-dependent scattering. Both particles can be probed in future DD experiments.




\bibliographystyle{aipproc}   



\end{document}